\documentclass{aa}
\usepackage[varg]{txfonts}
\usepackage{natbib}
\usepackage{graphicx}
\usepackage{times}
\usepackage{color}
\bibpunct{(}{)}{;}{a}{}{,}

\newcommand{\oca}{$o$~Cas }
\newcommand{\oce}{$o$~Cas}

\newcommand{\spefo}{{\tt SPEFO} }

\newcommand{\respefo}{{\tt reSPEFO} }
\newcommand{\respefoe}{{\tt reSPEFO}}

\newcommand{\fotel}{{\tt FOTEL} }

\newcommand{\korel}{{\tt KOREL} }

\newcommand{\m}{$^{\rm m}\!\!.$}
\newcommand{\ANG}{\accent'27A}

\newcommand{\ks}{km~s$^{-1}$}

\newcommand{\ms}{M$_{\odot}$}
\newcommand{\rs}{R$_{\odot}$}

\newcommand{\ha}{H$\alpha$ }

\newcommand{\mgii}{\ion{Mg}{ii}~4481~\ANG}

\newcommand{\n}{\hbox{\v{n}}}

\begin{document}

   \title{The Be star $o$ Cas is indeed the primary of a triple system}
  \author{P.~Harmanec\inst{1}\and
          S.~Yang\inst{2}\and
          M.~\v{S}lechta\inst{3}\and
          E.D.~Grundstrom\inst{4}\and
          J.~Ribeiro\inst{5}\and
          A.~Harmanec\inst{6}
}
   \offprints{P. Harmanec\,\\
               \email Petr.Harmanec@matfyz.cuni.cz}

  \institute{
   Astronomical Institute of Charles University,
   Faculty of Mathematics and Physics,\hfill\break
   V~Hole\v{s}ovi\v{c}k\'ach~2, CZ-180 00 Praha~8 - Troja, Czech Republic
 \and
 Department of Physics and Astronomy, University of Victoria,
 P.O. Box 1700, STN CSC, Victoria BC V8W 2Y2, Canada
 \and
   Astronomical Institute, Czech Academy of Sciences,
   CZ-251 65 Ond\v{r}ejov, Czech Republic
 \and
 Vanderbilt University, Nashville, Tennessee, USA
 \and
 Observatório Astronómico do Centro de Informação Geoespacial do Exército,
 Lisboa, Portugal
 \and
  Faculty of Mathematics and Physics, Charles University, Prague, Czech Republic
}
\date{Accepted April 15, 2026}

  \abstract{Analysis of radial velocities of two narrow absorption components
in the \ion{Mg}{ii}~4481~\AA\ line demonstrated that the secondary of
the Be star $o$~Cas is indeed a~close binary system composed of two B7 stars
orbiting each other with a period of 11\fd6604. Orbital solutions and spectral
disentangling lead to consistent system properties. The system is extremely
important for the research of Be stars since its future interferometric
observations with a~high spatial resolution could allow the mass
and perhaps even the radius of a Be star to be derived without too many model assumptions,
mainly on the dynamical grounds.}


\authorrunning{P. Harmanec et al.}
\titlerunning{$o$ Cas is a triple system}
\maketitle
\nolinenumbers

\section{Introduction}
The B2V - B5III-IVe star \oca (22~Cas, HD 4180, BD+47$^\circ$183, MWC~8, HIP~3504)
is the brighter component of a wide visual system WDS~J00447+4817, which exhibits
few or no signs of orbital motion. Component B is an 11\m2 F8 star
BD+47$^\circ$183B at recorded separations ranging from 32\farcs8 to 34\farcs4.
\citet{abt78} concluded that \oca is a single-line spectroscopic binary
with a period of 1033~days and a nearly circular orbit.
Although this conclusion was questioned in some studies, the
existence of such a pair was ultimately confirmed; see \citet{zarf28}
and references therein. The orbit is circular with a period
of $1031\fd55\pm0\fd71$ and a remarkably large semi-amplitude of more
than 20~\ks. This represents a problem, since for the inclination of the orbit
derived from interferometry, the companion to the Be primary should be
the more massive of the two. However, there is no trace of it in the optical spectra.
To explain this puzzle, \citet{zarf28} tentatively suggested that the companion
could actually be a close binary composed of two A stars. Inspired by this
suggestion, \citet{grundstrom2007} carefully inspected the Kitt Peak
National Observatory (KPNO) blue spectra at her disposal and discovered two weak
and narrow absorption lines in the core of the broad \mgii\ line, which had
been changing their positions on a timescale of days, perhaps with about
a four-day period.

Since no further studies of this possible binary appeared, this prompted us to
combine our efforts, obtain new spectra, and analyse them in an effort to prove
the existence of such a close binary and to derive its orbital elements.

\section{Spectroscopic data and their reductions}

Throughout this paper, we shall use the dates of observations expressed in
the `reduced heliocentric Julian date' (RJD) defined as

\centerline{RJD=HJD$-$2400000.0\,.}

\begin{table}
\begin{center}
\caption[]{Journal of blue electronic spectra of \oce.}\label{jourv}
\begin{tabular}{ccrcrl}
\hline\hline\noalign{\smallskip}
Spg.& Time interval  &No.&Wavelength &Spectral  \\
 No.&                &of & range     &res.\\
    &     (RJD)      &RVs& (\AA)   \\
\noalign{\smallskip}\hline\noalign{\smallskip}
 1&52695.29--52695.34& 2&4200--4750&40000\\
 2&52910.00--52911.98& 7&4460--4600&42600\\
 3&52910.38--52910.50& 3&4380--4630&11700\\
 4&52959.59--52981.97& 9&3870--4500&14200\\
 5&53233.92--55163.73&14&4230--4585$^*)$& 13000\\
 6&55801.57--59179.34&22&4750--6990&10000\\
 7&60114.55--60132.54& 6&4410--4540&51600\\
 8&60270.32--60344.33& 7&4380--4635&10600\\
 2&60523.79--60528.97&30&4380--4520&42600\\
 2&60586.67--60587.02& 4&4380--4520&42600\\
 2&60879.82--60881.97& 7&4380--4520&42600\\
 2&60921.77--60922.00& 6&4380--4520&42600\\
 2&60937.70--60938.04& 7&4380--4520&42600\\
 2&61060.65--61062.80&13&4380--4520&42600\\
\noalign{\smallskip}\hline\noalign{\smallskip}
\end{tabular}
        $^*)$ Four of the KPNO spectra cover the range from 4300 to 4650~\AA.
\tablefoot{Column ``Spg. No.": \ \
        1... OND 2.0 m reflector, Cassegrain Heros spectrograph \citep{heros};
2... DAO 1.2 m reflector, coud\'e grating McKellar spectrograph, 9682M configuration,
     CCD SITe4 detector;
3... OND 2.0 m reflector, coud\'e spectrograph, CCD SITe5 2030x800 pixel detector;
4... DAO 1.2 m reflector, coud\'e grating McKellar spectrograph, 32121B configuration,
     CCD SITe4 detector;
5... KPNO 0.9~m reflector, coud\'e feed spectrograph;
6... BeSS amateur echelle spectra (\citealt{neiner2011};
\url{http://basebe.obspm.fr/basebe});
7... OND 2.0 m reflector, OES echelle spectrograph, CCD EEV 42-40-1-368 detector;
8... Lisbon Celestron C14 0.356~m reflector, LhiresIII spectrograph.
}
\end{center}
\end{table}

\noindent Inspecting the Ond\v{r}ejov (OND) and the Dominion Astrophysical Observatory
(DAO) data archive, we actually found several pre-discovery blue spectra
of \oce. New electronic spectra were obtained at KPNO,
DAO, OND, and Lisbon. We also used 22 amateur spectra from the BeSS database
\citep{neiner2011} having a spectral resolution of 10000 or better.
The journal of all spectroscopic observations used here is in Table~\ref{jourv}.

   The initial reduction of all spectra (bias subtraction, flatfielding,
and creation of 1D frames) was carried out with the pipelines used
in individual observatories. Further reduction (spectra normalisation,
cleaning from cosmics and flaws, and radial-velocity (RV) measurements)
were carried out in the programme
\respefo2. This programme is written in JAVA and can run on different
platforms (Linux, Windows, MacOS). It is being developed by Adam~Harmanec.
It can, among other things, import spectra that were originally reduced
in \spefo \citep{spefo,spefo3} and treat spectra stored as FITS files.
For the DAO spectra, it also provides wavelength calibration.
The programme is described in more detail in Appendix~\ref{ape}.

\begin{figure}
\resizebox{\hsize}{!}{\includegraphics[angle=0]{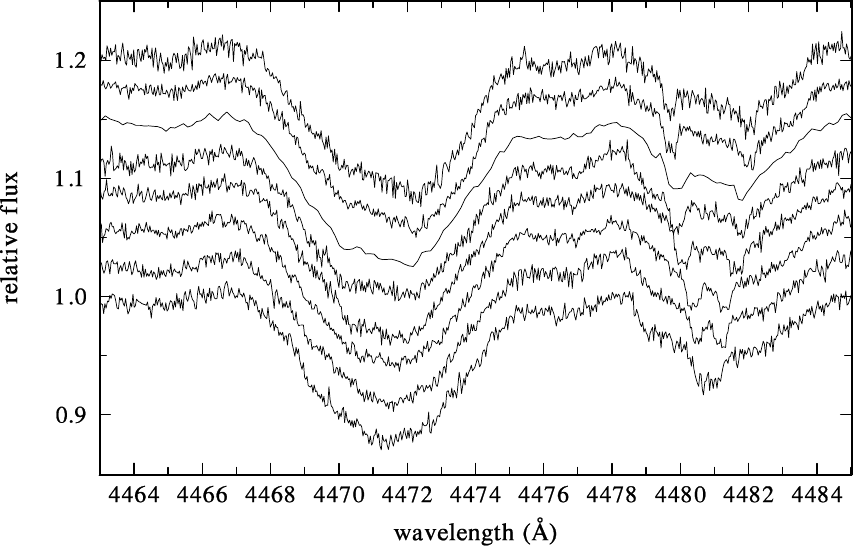}}
\caption{A pre-discovery series of one OND and seven DAO spectra in the
neighbourhood of the \ion{He}{i}~4471.508 and \ion{Mg}{ii}~4481.228~\AA\
from September 2003. From top to bottom the RJDs of the spectra are
52909.9993, 52910.0102, ...10.4959 (OND), ...10.6240, ...10.9934,
52911.6137, ...11.7928, and ...11.9786. A smooth change in the position
of the two narrow \ion{Mg}{II} lines in time is clearly seen.}\label{mga}
\end{figure}

\begin{figure}[t]
\resizebox{\hsize}{!}{\includegraphics[angle=0]{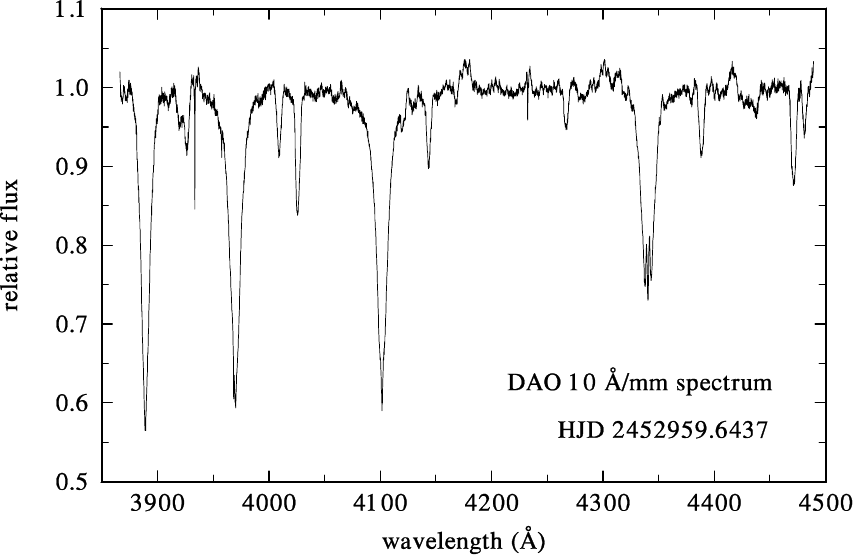}}
\resizebox{\hsize}{!}{\includegraphics[angle=0]{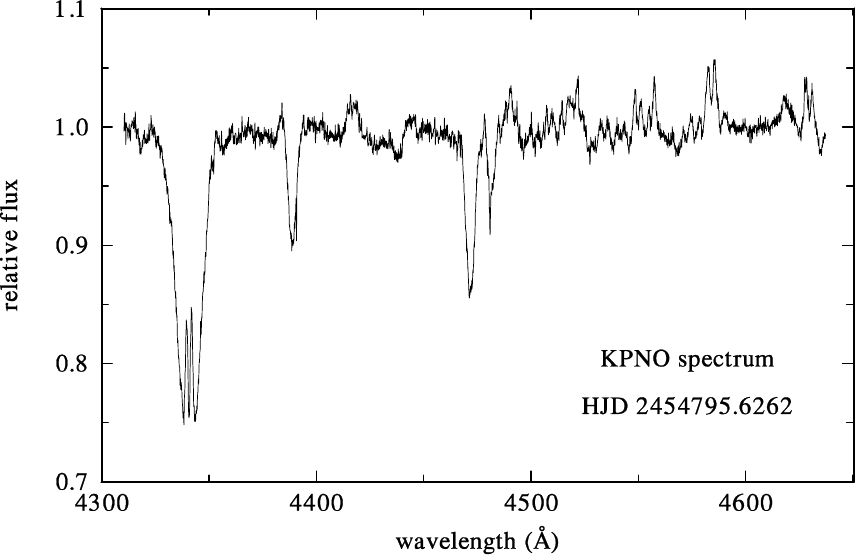}}
\caption{Examples of some blue spectra at our disposal.}\label{bluesp}
\end{figure}

Two examples of the blue spectra that we have are in Fig.~\ref{bluesp}.
It is seen that many weak metallic lines in the emission are present, which
makes the task of proper spectra rectification quite challenging.

\begin{figure}[t]
\resizebox{\hsize}{!}{\includegraphics[angle=0]{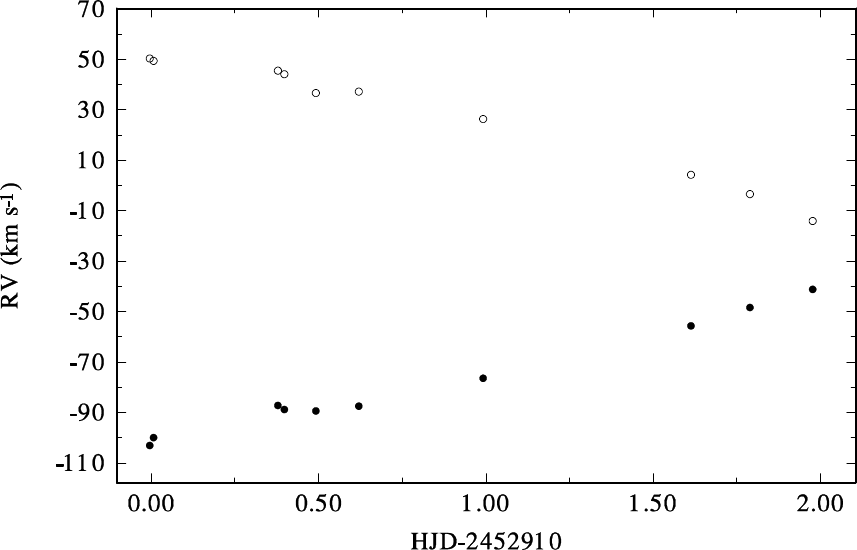}}
\resizebox{\hsize}{!}{\includegraphics[angle=0]{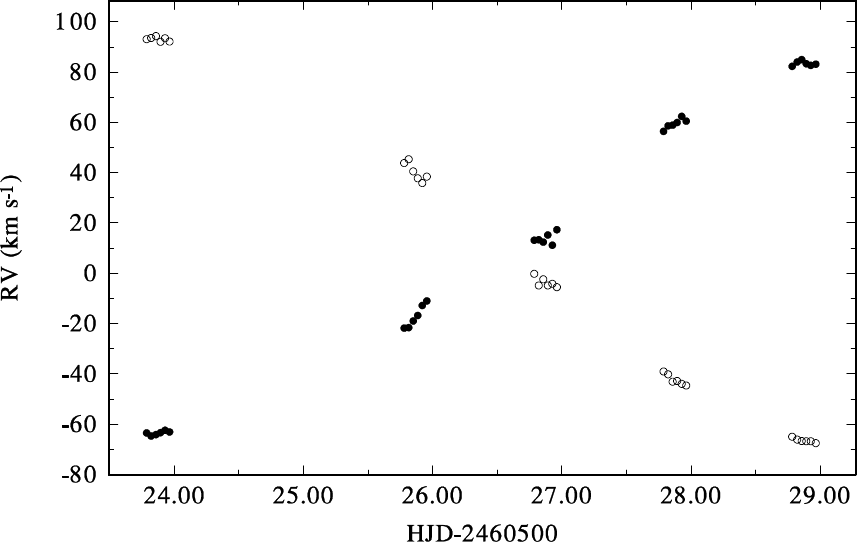}}
\resizebox{\hsize}{!}{\includegraphics[angle=0]{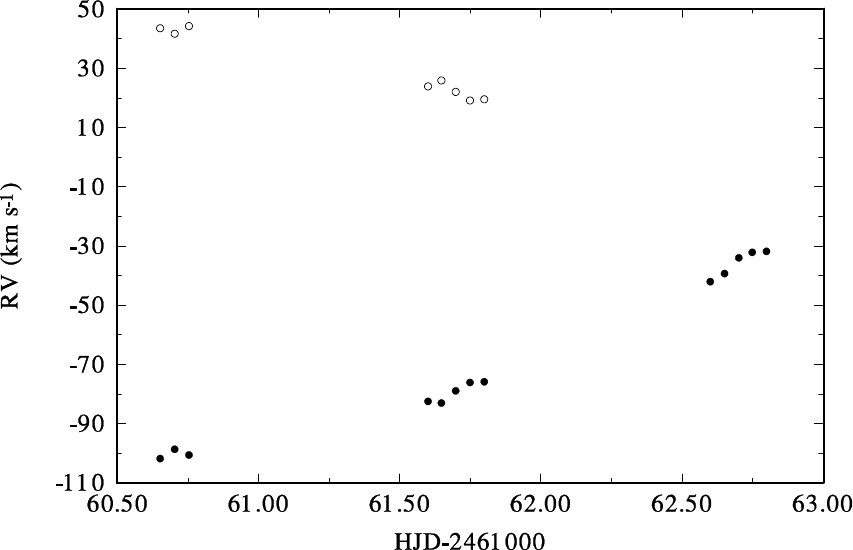}}
\caption{Comparison of the RVs of the two narrow components
of the \ion{Mg}{ii}~4481.228~\AA\ line for three available
time segments. The RVs of the primary and secondary are shown
by black dots and open circles, respectively.}\label{rvtime}
\end{figure}

\section{The inner orbit}
\begin{figure}[t]
\resizebox{\hsize}{!}{\includegraphics[angle=0]{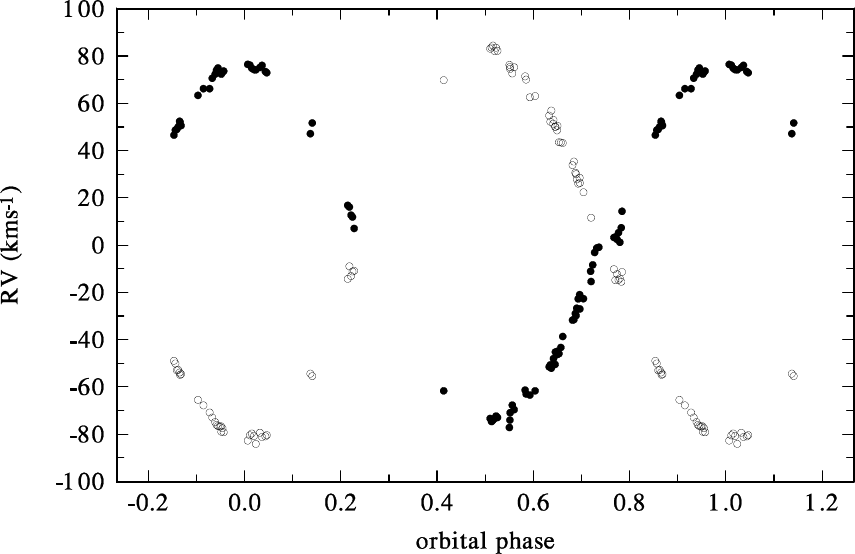}}
\resizebox{\hsize}{!}{\includegraphics[angle=0]{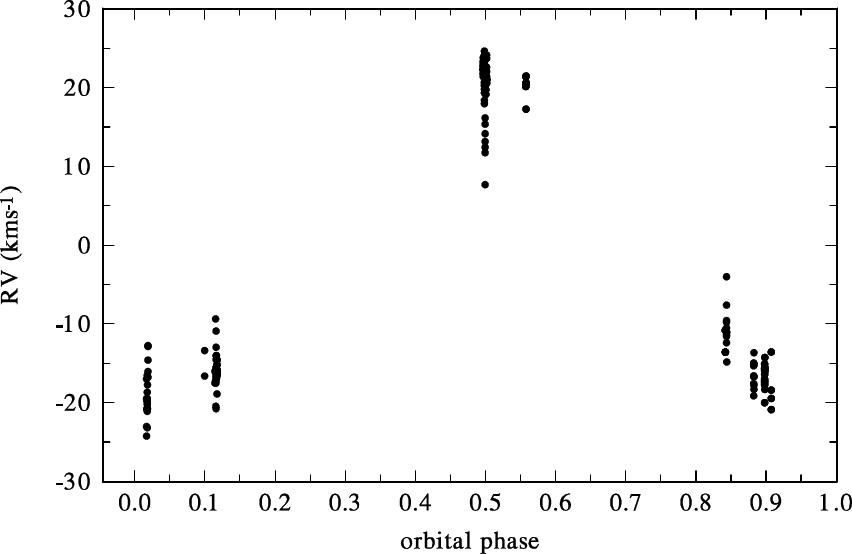}}
\caption{Top: RV curves of the narrow \ion{Mg}{ii} lines for the period of
11\fd066043(31) based on a \fotel solution for RVs measured in \respefoe.
Bottom: RV curve of the short-period binary around the centre of mass of
        the 1031\fd55 binary based on the same RVs.}\label{orbit}
\end{figure}

In spite of the fact that we managed to accumulate 85 higher-resolution spectra
with the \ion{Mg}{ii}~4481.228~\AA\ sharp lines well visible (out of the total
number of 137 spectra listed in Table~\ref{jourv}), finding the true orbital period
of the inner orbit was all but easy. The two weak and narrow lines are rather similar
to each other and it is easy to misinterpret one for another. Another complication
arises from the fact that the whole binary moves with the Be tertiary in
the 1032~d orbit with a non-negligible semi-amplitude of about 20 \ks.
It was necessary to obtain spectral series covering several consecutive days
to see the relative motion of the two components in continuous time sequences.
Three limited series of spectra obtained within about a week of consecutive
observations were obtained. Their RVs for (seemingly) stronger and weaker components of
narrow \ion{Mg}{ii} lines were plotted versus time in the three panels of
Fig.~\ref{rvtime}.  After a number of trials for period searches, separately for the RVs
of the stronger and fainter components, using a programme based on the \citet{deeming75}
period search technique, we concluded that the most probable
orbital period of the inner system is close to 11\fd7. Then we used the programme \fotel
\citep{fotel1, fotel2}, which is able to derive orbital solutions for triple systems.
In our case, we kept the orbital elements of the outer 1031\fd55 orbit fixed at values
obtained by \citet{zarf28} from a very long series of observations. After properly identifying the
primary and secondary components, we finally arrived at a very satisfactory
solution presented in Table~\ref{sol}. The phase plots of all well-resolved RVs
measured in \respefo through a comparison of direct and flipped line-profile images
are in Fig.~\ref{orbit}.

\begin{table}
\begin{flushleft}
\caption{Orbital elements based on \respefo RVs of the
        two narrow \ion{Mg}{ii}~4481~\AA\ lines and a \fotel
        triple-star solution.}
\label{sol}
\begin{tabular}{lccrrccc}
\hline\hline\noalign{\smallskip}
        Element                   & \fotel           & \korel        \\
\noalign{\smallskip}\hline\noalign{\smallskip}
$P$ (d)                    & $11.66043\pm0.000031$   & 11.66045  \\
$T_{\rm RV max.1}$ (RJD)   & $60529.5023\pm0.0078$   & 60529.5332\\
$e$                       &   0.0 fixed              & 0.0 fixed \\
$\gamma$ (\ks)             &$-10.65\pm0.22$          & -10.5     \\
$K_1$    (\ks)             & $75.79\pm0.45$          & 78.58     \\
$K_1/K_2$                  &$0.9303\pm0.0073$        & 0.9675    \\
$K_2$    (\ks)             & 81.47                   & 81.22     \\
$K_3/K_1$                  & $0.2703\pm0.0038$       & 0.2695    \\
$K_{1+2}$ (\ks)            & 20.487                  & 21.18     \\
$K_3$     (\ks)            & 21.000                  & 23.05     \\
$m_1\sin^3i$ (\ms)         & 2.434                   & 2.506  \\
$m_2\sin^3i$ (\ms)         & 2.264                   & 2.425  \\
$a\,\sin i$ (\rs)          & 36.24                   & 36.83  \\
No. of RVs (prim./sec.)    & 85/80                   & 137 \\
rms (\ks)                  &  2.76                   & --  \\
\hline\noalign{\smallskip}
\end{tabular}
\end{flushleft}
\end{table}

\begin{figure}[t]
\resizebox{\hsize}{!}{\includegraphics[angle=0]{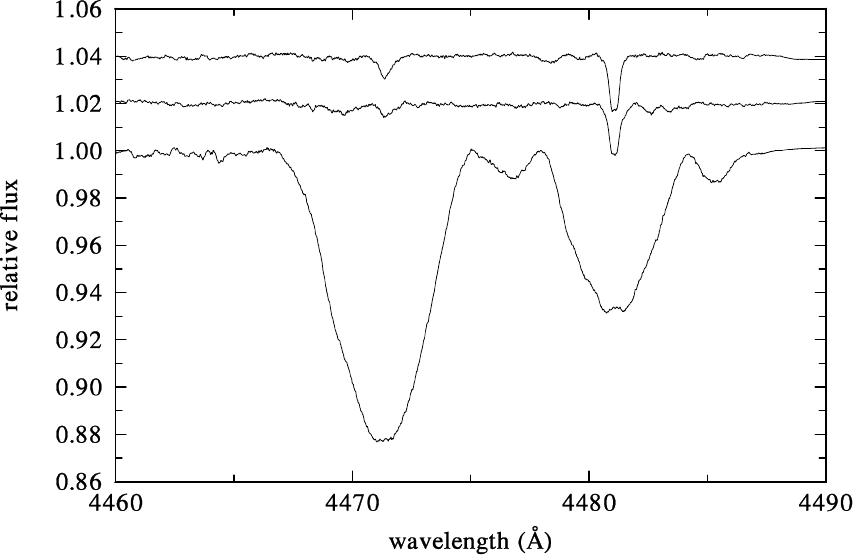}}
\caption{The disentangled line profiles of the \ion{He}{i}~4471~\AA,
and \ion{Mg}{ii}~4481~\AA\ line for the three components of the triple
system. All profiles are normalised to the joint continuum of the whole
system and the profiles of the primary and secondary of the 11\fd66 subsystem
were shifted for 0.04, and 0.02, respectively, in relative flux.}
\label{korprof}
\end{figure}

To confirm the result, we then used all 137 available spectra
to disentangle them with the programme \korel \citep{korel1,korel2,korel3}.
The resulting orbital elements are also listed in Table~\ref{sol} and
the disentangled line profiles of all three stars are shown in
Fig.~\ref{korprof}. The referee called our attention to the fact
that the disentangled \ion{Mg}{ii} line profile of the Be primary seems
to have a small central peak, reminiscent of central quasi-emission peaks
found for several Be stars seen roughly equator-on
\citep[see][and references therein]{rivi99}. Although we agree that this
possibility deserves further investigation with high-quality spectra,
for the moment we conclude that the effect is only an artefact of the
disentangling process. The same feature is not seen in the neighbouring stronger
\ion{He}{i} line, and we have verified that it is absent if we
disentangle only the high-resolution spectra.

\section{Probable properties of the triple system}
Obviously unaware of the \citet{zarf28} study, \citet{videla2022} and
\citet{angu2023} attempted to estimate the masses of the wide 1031\fd55 binary system \oca
using Bayesian inference. \citet{videla2022} estimated the
mass of the Be primary as ($5.62\pm1.19$)~\ms\, while \citet{angu2023}
derived masses as 5.77 and 4.66~\ms, however with huge errors.

 Our study allows us to provide more realistic estimates of the basic
physical properties of the system and its components. Based on the \korel solution
of Table~\ref{sol} and assuming that the inner system is observed
under the same orbital inclination of $65^\circ$ as that estimated from
interferometry for the outer system, we obtain the masses
of the components of the inner binary, 3.37~\ms\ and 3.26~\ms, therefore
the total mass of the inner binary is 6.63~\ms. From the solution
for the outer system, we obtain the total mass of the inner binary as
6.73~\ms, the mass of the distant Be component being 6.07~\ms. These
two results are in very good agreement. We note that according
to the tabulation of normal masses by \citet{mr88} the two components
of the inner binary are probably B7 stars, while the Be component would
correspond to a normal star of spectral type B3, in accordance with available
spectral classifications. We wish to point out that \oca is an astrophysically
unique and very important system. Its future interferometric observations
with a~high spatial resolution should permit the first determination
of the precise mass of a Be star based solely on the dynamical considerations,
without the use of any statistical relations between the spectral type and
stellar properties. A very favourable circumstance is also the relatively
short outer orbit, with semi-amplitudes of the orbital motion of about 20~\ks,
which guarantee good accuracy of mass determination.

It is true that dynamically determined masses of several Be stars in binaries and/or
multiple systems have already been published; see, for example \citet{klement2022}
and references therein. However, some of these estimates depend either
on distance estimates or on less certain RV curves of compact
companions detected in the far-UV spectra.

  We are aware of only the following few triple systems
containing a Be star; all are much less favourable for
 accurate mass determination than \oce. For V2048~Oph = 66~Oph, an~inner binary
rather similar to \oce, composed of two late B or early A stars orbiting
each other with a period of 10\fd78, was found by \citet{stefl2004}.
However, the outer orbit of this binary with the Be primary is very
long, $23421\fd1\pm4\fd1$ according to \citet{Hutter2021}, and not very
favourable for mass determination.
Another triple system, $\nu$~Gem = HD~45542 \citep{klement2021} consists of two stars
orbiting with a period of 53\fd8 and a distant Be star with a long outer
orbit of about 7000~d. The \ha emission of the Be star is rather weak,
and the RV curve of the Be star was thus derived from the RV of the \ha shell
absorption line, which does not need to return the true RV amplitude.
V1371~Tau = HD~36665 \citep{rocha2026} consists of three early B stars. Two of them
form a peculiar eccentric-orbit eclipsing binary with a period of 33\fd62
and the third body is a Be star in uncertain $\sim11$~yr period. There is
a suspicion of secular change of the orbital inclination of the inner
eclipsing binary, and there are also rapid light changes present in the system.
A~well-known B6e star $o$~And = HD~217675 is in a~wide orbit of a still
uncertain period of several years with two sharp-lined stars, which revolve
around each other with a~33\fd0 period \citep{hill88,zhuchkov2010}.

  In passing, we note that similar triple or even multiple systems represent
an interesting challenge to the theory of their evolutionary history. In
addition to \oca and the systems discussed above, one can also mention the compact triple
system $\xi$~Tau \citep{xitau2016}, with a rapidly rotating B star, not known
to have Balmer emission.

\begin{acknowledgements}
We gratefully acknowledge the use of the latest publicly available versions
of the programmes \fotel and \korel written by P.~Hadrava.
We thank Marek Skarka who obtained two of the OND OES spectra in service mode
for us. Several constructive suggestions by an~anonymous referee
helped to improve the paper and are gratefully acknowledged.
Based on spectroscopic CCD observations obtained at the Dominion Astrophysical
Observatory, Herzberg Astronomy and Astrophysics Research Centre, National Research
Council of Canada, on new and archival spectroscopic observations obtained with
the Heros echelle spectrograph, linear CCD coud\'e-focus spectrograph, and
Ond\v{r}ejov echelle spectrograph, all three attached to the 2.0~m reflector of
the Astronomical Institute of Czech Academy of Sciences, on CCD spectra from
the coud\'e feed spectrograph attached to the Kitt Peak 0.9~m reflector,
on CCD spectra from Lisbon 0.356~m reflector, and on amateur echelle spectra
from the BeSS database, operated at LESIA,
Observatoire de Meudon, France: \url{http://basebe.obspm.fr}.
The collection of spectra was supported by the grants
205/06/0304, 205/08/H005, P209/10/0715, and GA15-02112S of the Czech
Science Foundation.
Finally, we acknowledge the use of the electronic database from
the CDS, Strasbourg, and the electronic bibliography maintained by
the NASA/ADS system.
\end{acknowledgements}

\bibliographystyle{aa}
\bibliography{zarfin}

\begin{appendix}
\nolinenumbers
\section{Program \respefo for the reductions and measurements of 1D electronic spectra}
\label{ape}

Program \respefoe, written in Java by Adam Harmanec, is a modern replacement for
the original \spefo program developed in Pascal by the late Ji\v{r}\'\i\ Horn \citep{spefo}.
It runs on different platforms (Unix, Windows, macOS) and provides a comprehensive
environment for one-dimensional spectral analysis. \respefo operates primarily
on 1D spectra produced by standard observatory pipelines. For the majority of observatories,
the spectra in 1D frames are already stored as pairs of wavelength and relative flux,
typically in FITS format. The program converts imported data into an internal {\tt .spf}
format, which preserves raw data, metadata, preprocessing steps, and measurement results,
allowing workflows to be revisited without repeating earlier steps. Besides spectra recorded
in the standard FITS format, the current version 2 of \respefo can process spectra from
the CHIRON echelle spectrograph from CTIO, spectra from the BeSS database, FEROS echelle
spectra, spectra from the Hercules echelle spectrograph, and also spectra from the DAO,
for which it can also derive the wavelength scale from measurements of comparison
ThAr or FeAr spectra. It is also able to import spectra reduced earlier with the original
\spefo as well as plain ASCII files recorded as wavelength - relative flux pairs.

The usual sequence of reduction steps is to define the project, import the spectra,
rectify them, clean them of cosmics and residual flaws not removed by standard observatory
pipelines, and measure RVs and spectrophotometric quantities of selected spectral lines.
The tracing paper method of RV measurement allows flexible settings on different parts of
more complicated line profiles in spectra from several components of a multiple stellar
system. For red and infrared spectra, it is also possible to measure a selection of
telluric lines to apply small additional corrections to the zero point to bring spectra
from different instruments onto a common wavelength scale. The tracing paper method is
illustrated in Appendix~C of the paper by \citet{hec2020} and its practical realisation
in the program \respefo is described in detail in Sect.~3.1 of the paper
by \citet{wolf2021}.

Version~2 of the program, together with a detailed documentation and user manual
can be downloaded at \url{https://astro.troja.mff.cuni.cz/projects/respefo}, separately
for Unix, Windows, and macOS operating systems.

The software is distributed under the EPL~2.0 license and remains
under active development.

\end{appendix}
\end{document}